\title{Analysis of Distributed Snapshot Algorithms}
\author{
Sharath Srivatsa\\
       sharath.srivatsa@iiitb.org
 }
\date{\today}

\documentclass[12pt]{article}
\usepackage{booktabs}
\usepackage{graphicx}
\graphicspath{{./}}
\begin{document}
\maketitle

\begin{abstract}
Many problems in distributed systems can be cast in terms of the problem of detecting global states. For instance, the global state detection algorithm helps to solve an important class of problems: stable property detection. A stable property is one that persists: once a stable property becomes true it remains true thereafter. Examples of stable properties are ``computation has terminate'', ``the system is deadlocked'' and ``all tokens in a token ring have disappeared''.  \cite{cl:seminal}

Distributed Snapshot algorithms are categorized by underlying message delivery mechanisms FIFO, Non-FIFO and Causal Ordering. Through FIFO channels the messages arrive in the order in which they were transmitted and in Non-FIFO channels the order is not ensured. Causal Ordering mechanism delivers the messages in the order they were created. 

Snapshot recording durations at each process contribute to the overall efficiency of the algorithm. In this paper we are presenting the observed variations in snapshot recording durations at processes in a distributed system.  We conclude with key characteristics of a reliable and effective snapshot algorithm. Simulations were achieved using SimGrid Java API.
\end{abstract}

\section{Introduction}
A snapshot of a distributed system is a global state (consisting of the local states of the processes and all the messages in transit) which is meaningful in the sense that it corresponds to a possible global state where the local states of all processes and of all communication channels are recorded simultaneously 	\cite{mat:vclks}. Since processes do not share common clock, the start and end of recordings varies at each process. Through simulations we have captured recording durations at each process in a distributed system. 

Algorithms in Table 1 were reviewed and analysed.

Chandy-Lamport algorithm assumes FIFO channels and rely on control messages. Spezialetti-Kearns algorithm optimizes concurrent initiation of snapshot collection and efficiently distributes the recorded snapshot with channel recording similar to Chandy-Lamport.

Lai-Yang algorithm assumes non-FIFO channels and does not require control messages since colouring scheme is used on computation messages. Initially all messages are ``white'' and ``red'' message initiates the snapshot recording. Mattern algorithm initiates snapshot recording through vector clocks.

Acharya-Badrinath and Alagar-Venkatesan algorithms rely on underlying causal delivery of messages and use control messages to capture snapshot.

Distributed system was simulated using SimGrid platform with 90 hosts. Latency and message interval variations were generated using Poisson, Pareto, Weibull and ARIMA distributions. In an instance of simulation 10 messages were sent by each process and snapshot was initiated once. Snapshot recordings of processes were not aggregated, instead the recording durations at each process were captured and analysed.

\begin{table}[!htb]
\centering
\caption{Message delivery mechanisms and distributed snapshot algorithms}
\begin{tabular}{ccc}
\toprule FIFO & Non-FIFO & Causal Delivery\\
\midrule Chandy-Lamport \cite{cl:seminal} & Lai-Yang \cite{ly:coloring} & Acharya-Badrinath \cite{ab:causal}\\ Spezialetti-Kearns \cite{sk:optcl} & Mattern \cite{mat:vclks} & Alagar-Venkatesan \cite{av:causal}\\
\bottomrule
\end{tabular}
\end{table}

\section{Simulation results and analysis}

The main work accomplished was running the snapshot algorithms with variations in message generation and platform latency from standard distributions. Chandy-Lamport, Lai-Yang and Mattern algorithms were implemented and analysed.  Acharya-Badrinath and Alagar-Venkatesan were similar, hence a combined implementation was done and analysed. Birman-Schiper-Stephenson Protocol was implemented to achieve causal ordering of messages. The snapshot recording durations at each process were considered to be the key factors to gain insights of the algorithms. A minimal variation in recording durations across processes would indicate better time complexity, though the space complexity be high.
\subsection{Typical platform}

The first simulation was achieved using distributed platform with multiple links of different latencies and the message generation interval from a normal distribution. Any two hosts were connected with varying number of links and every host was connected to other host with a set of links. Different processing power was assigned to each host. The duration at each host to record the snapshot was platform specific time units. Recording durations at each host are of the same units.
Figure 1 shows the variations in the algorithms.

It can be observed that there are more variations in Mattern and Lai-Yang algorithms, this is because control messages are not used. Less variations in Chandy-Lamport and Acharya Badrinath-Alagar Venkatesan algorithms are due to use of control messages. Control messages ensure that the algorithm completes in a minimal time.

We further verified the variations in recording durations using only varying latencies from standard distributions and, using both varying latencies from standard distributions and Poisson message intervals.

\begin{figure}[!htb]
\centering
\includegraphics[height=3in, width=3in]{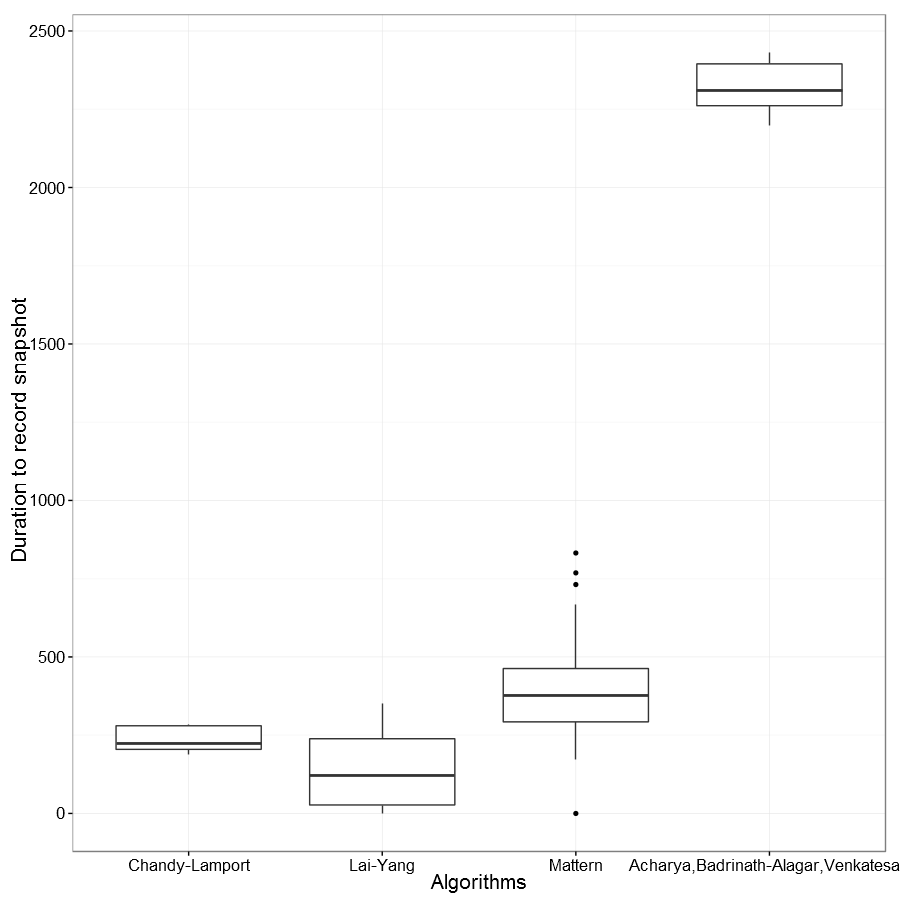}
\caption{Recording duration variations on a typical platform}
\vskip -6pt
\end{figure}
\subsection{Platform with varying latencies}

Figure 2 shows the recording durations with latencies from Poisson, Pareto, Weibull and ARIMA distributions.

Figure 3 shows the recording durations with Poisson message interval and latencies from Poisson, Pareto, Weibull and ARIMA distributions.

Table 2 and 3 shows the standard deviations of recording durations.

\begin{figure*}[!htb]
\centering
\includegraphics[height=4in, width=4in]{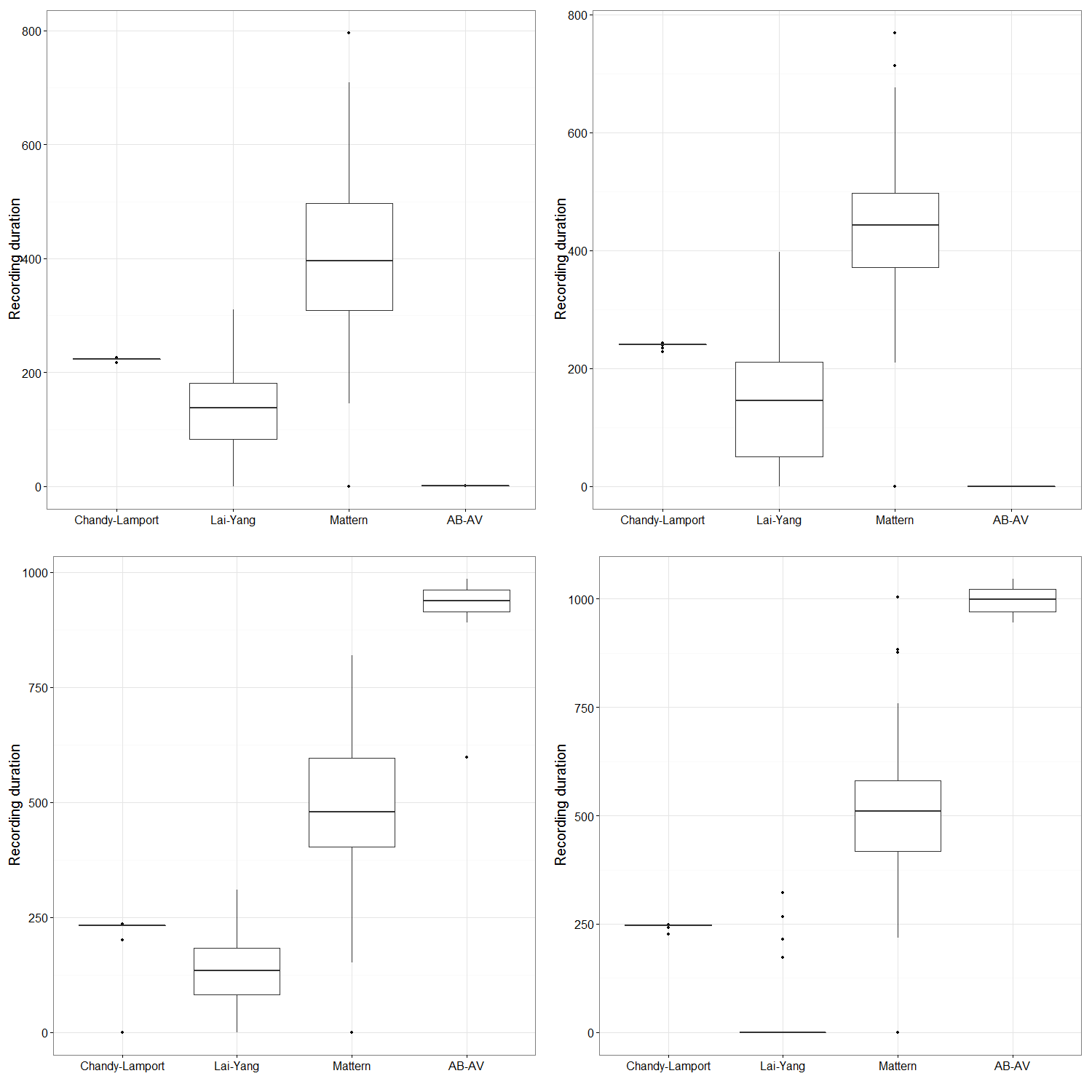}
\caption{Recording duration variations with varying link latencies}
\end{figure*}

\begin{table*}[!htb]
\centering
\caption{Standard Deviations of recording durations with varying latencies}
\begin{tabular}{ccccp{4cm}}
\toprule Latency Distribution & Chandy-Lamport & Lai-Yang & Mattern & Acharya Badrinath - Alagar Venkatesan\\
\midrule Possion & 0.804 & 75.176 & 135.666 & 0.001 \\ Pareto & 97.206 & 97.206 & 107.792 & 0 \\ Weibull & 3.429 & 72.711 & 132.18 & 27.639 \\ ARIMA & 1.424 & 422.13 & 1571.357 & 27.45 \\
\bottomrule \end{tabular} 
\end{table*}

\begin{figure*}[!htb]
\centering
\includegraphics[height=4in, width=4in]{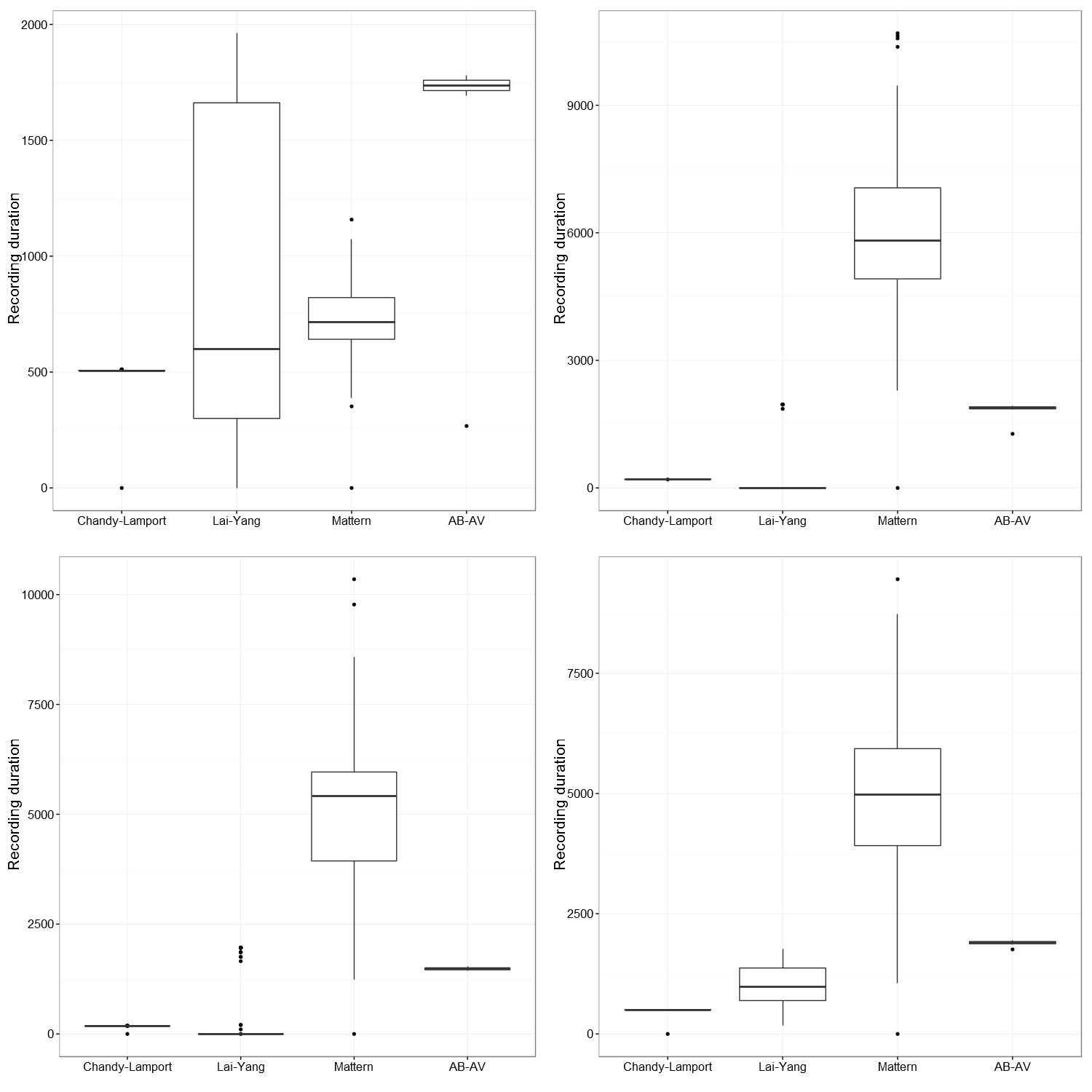}
\caption{Recording duration variations with Poisson message interval and varying link latencies}
\end{figure*}

\begin{table*}[!htb]
\centering
\caption{Standard Deviations of recording durations with Poisson message intervals and varying latencies}
\begin{tabular}{ccccp{4cm}}
\toprule Latency Distribution & Chandy-Lamport & Lai-Yang & Mattern & Acharya Badrinath - Alagar Venkatesan\\
\midrule Possion & 1.323 & 715.236 & 159.897 & 25.837 \\ Pareto & 0.783 & 490.815 & 1755.357 & 27.289 \\ Weibull & 0.454 & 733.017 & 1685.168 & 26.872 \\ ARIMA & 2.325 & 52.245 & 145.374 & 29.682 \\
\bottomrule \end{tabular} 
\end{table*}

It can be observed that even with random latencies, algorithms (Chandy-Lamport and Acharya Badrinath-Alagar Venkatesan) have minimal recording variations across hosts. In these algorithms control messages bound the duration to complete snapshot recording. Whereas in Lai-Yang and Mattern algorithms the snapshot initiation is piggybacked on the computation message, hence the snapshot recording is dependent on computational messages which is dependent on the system rather on the algorithm. 

Minimal standard deviation indicates high probability of better time complexity. Control messages trigger the snapshot recording at each host in a bounded manner.

\section{Conclusions}
Through simulations we observed that recording durations were minimal when control messages are used and causal message delivery ensured reliability. Control messages bound the completion time of the algorithm. We conclude that a snapshot algorithm using causal delivery and control messages sent with vector clock timestamp would be an effective solution.

\bibliographystyle{abbrv}
\bibliography{Simple_Format}

\end{document}